\documentclass[twocolumn,showpacs,amsmath,amssymb,prb,graphics,graphicx,superscriptaddress
]{revtex4-1}
\usepackage{bm}
\usepackage[pdftex]{graphicx}
\usepackage{dcolumn}
\begin{document}

\title{London penetration depth at zero temperature and near superconducting transition}

\author{V. G. Kogan}
\email{kogan@ameslab.gov}

 \affiliation{Ames Laboratory, Ames, IA 50011, USA}
 
\author{M. A. Tanatar}

\email{tanatar@ameslab.gov}
 \affiliation{Ames Laboratory, Ames, IA 50011, USA}
 \affiliation{Department of Physics and Astronomy, Iowa State University, Ames, IA 50011, USA}
 
\author{R. Prozorov}

\email{prozorov@ameslab.gov}
 \affiliation{Ames Laboratory, Ames, IA 50011, USA}
 \affiliation{Department of Physics and Astronomy, Iowa State University, Ames, IA 50011, USA}
 
 \begin{abstract}
A simple  relation is established between the   zero-$T$ penetration depth $\lambda (0)$ and the slope of $\lambda^{-2}(T)$ near $T_c$, similar to Helfand-Werthamer's relation for $H_{c2}(0)$ and the slope of $H_{c2}(T)$ at $T_c$  for the isotropic s-wave case with  non-magnetic scattering.\cite{HW} When the scattering parameter $\rho=\hbar v/2\pi T_c\ell$ ($v$ is the Fermi velocity and $\ell$ is the mean-free path) varies from 1 to 10, the coefficient of proportionality between   $\lambda^{-2}(0)$ and $T_c (d\lambda^{-2}/dT)_{T_c}$ changes from 0.43 to 0.38. Combining this relation with the Rutgers thermodynamic identity, one can express $\lambda (0)$ in terms of the slope $(dH_{c2}/dT)_{T_c}$ and the density of states.
\end{abstract}
\pacs{74.20.-z,74.76.-w,74.50.+r,85.25.Cp}
 \date{ \today}
\maketitle

\section{Introduction}

In the seminal work by Helfand and Werthamer   a simple relation between zero-temperature upper critical field $H_{c2}(0)$ and the slope of $H_{c2}(T)$ at $T_c$  was established for the isotropic s-wave superconductors with  non-magnetic scattering:\cite{HW}
 \begin{equation}
 H_{c2}(0) = \mu_{h}T_c\left(\frac{d H_{c2} }{dT}\right )_{T_c} \,.
 \label{HW0}
\end{equation}
With increasing transport scattering the numerical factor $\mu_h$ varies from 0.73 in the clean limit to 0.69 in the dirty case, i.e. $\mu_h\approx 0.7$ with $\sim 4$\% accuracy. This relation is commonly used to estimate  $H_{c2}(0)$ when only high temperature data for $H_{c2}(T)$ are available. It is shown below that a similar   relation exists for the penetration depth $\lambda(T)$:
 \begin{equation}
 \lambda ^{-2}(0)=\mu_\lambda T_c\left(\frac{d \lambda ^{-2}}{dT}\right )_{T_c} \,,
 \label{HW}
\end{equation}
with $\mu_\lambda\approx 0.4$ in a  range of scattering parameters covering nearly all practical transport scattering rates.

\section{  $\bm{\lambda^{-2}(0)}$ in terms of $\bm{T_c (d\lambda/dT)_{T_c}}$ }

The penetration depth in  isotropic BCS superconductors  is given by
 \begin{equation}
 \lambda ^{-2}= \frac{16\pi^2 e^2TN(0)\Delta ^2v^2
 }{3c^2}\,  \sum_{\omega>0}  \frac{1
}{\beta ^{2}\beta^\prime}  \,.
 \label{lambda-tensor1}
\end{equation}
Here, $\hbar\omega=\pi T(2n+1)$ defines the Matsubara frequencies, $N(0)$ is the one-spin density of states  at the Fermi level, $v$ is the Fermi velocity, $\Delta(T)$ is the gap parameter, $\beta^2 = \hbar^2\omega^2+\Delta^2 $, $\beta^\prime=\beta+\hbar/2\tau$, and $\tau$ is the transport scattering time. One can find this result in the book
by Abrikosov, Gor'kov and Dzyaloshinskii. \cite{3authors}  It can also be  derived   using Eilenberger quasi-classical version of the BCS theory, see, e.g., Refs.\,\onlinecite{nonloc,PK-ROPP}.

At zero temperature, one  replaces the sum with an integral according to $2\pi T\sum_\omega \to \int_0^\infty d(\hbar\omega)$ to obtain:\cite{Homes}
 \begin{eqnarray}
 \lambda ^{-2}(0)= \frac{4\pi^2  e^2N(0)
v^2}{3c^2\eta} \left(1 +\frac{4\tan^{-1}\frac{ \eta-1 }{\sqrt{1-\eta^2} }}{\pi\sqrt{1-\eta^2}}\right)
 ,\qquad
\label{lambda0}
\end{eqnarray}
where the scattering parameter
\begin{equation}
\eta=\frac{\hbar}{2\tau\Delta_0} = \frac{\pi}{2}\,\frac{\xi_0}{\ell},
\label{eta}
\end{equation}
 $\xi_0=\hbar v/\pi\Delta_0$ is the BCS zero-$T$ coherence length and   $\ell$ is the transport mean-free path.
  Eq.\,(\ref{lambda0}) works for any $\eta>0$. For $\eta>1$, it can be written in explicitly real form by replacing $\tan^{-1}\to -\tanh^{-1}$ and $\sqrt{1-\eta^2} \to \sqrt{ \eta^2-1}$.

Near $T_c$,   $\Delta^2=8\pi^2T_c(T_c-T)/7\zeta(3)$, $\beta\approx \hbar\omega =\pi T_c( 2n+1)$,   $\beta^\prime\approx    \pi T_c( 2n+1 +\rho )$ with
\begin{equation}
\rho=\frac{\hbar v}{2\pi T_c \ell}=e^{-\gamma}\eta ,
\label{rho}
\end{equation}
 $\gamma\approx 0.577$ is the Euler constant.
The sum in Eq.\,(\ref{lambda-tensor1}) is expressed in terms of digamma functions $\psi$. Doing the algebra one obtains the slope at $T_c$:
 \begin{equation}
\frac{d \lambda ^{-2}}{dT}=  \frac{64\pi e^2   N(0) v^2
 }{21\zeta(3)c^2T_c}\,   \frac{1}{\rho^2}\left[   \psi\left(\frac{1+\rho}{2}\right)-  \psi\left(\frac{1}{2}\right) -\frac{\pi^2\rho}{4}\right].
 \label{lambda-GL}
\end{equation}

\begin{figure}[h]
\includegraphics[width=7cm] {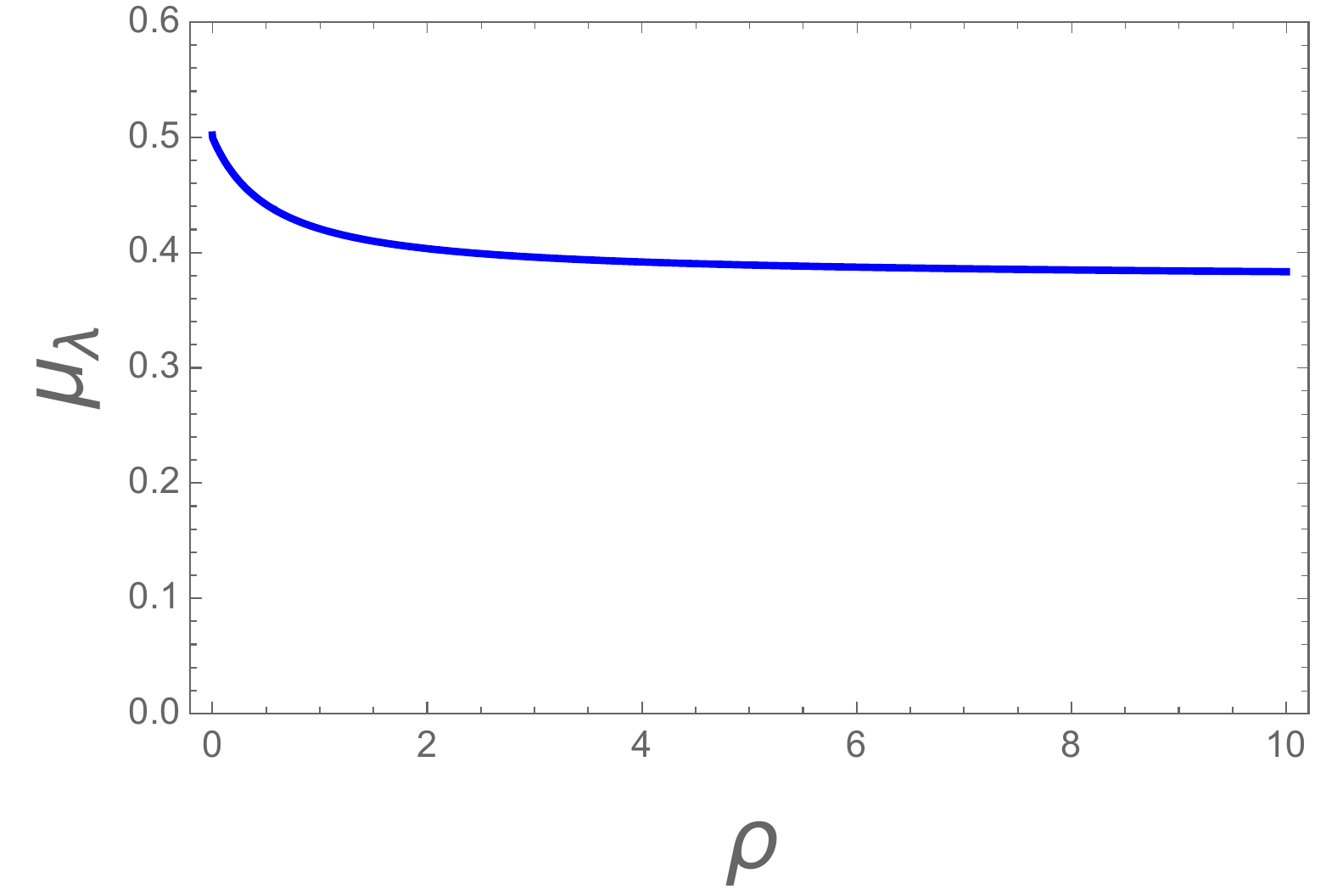}
\includegraphics[width=7cm] {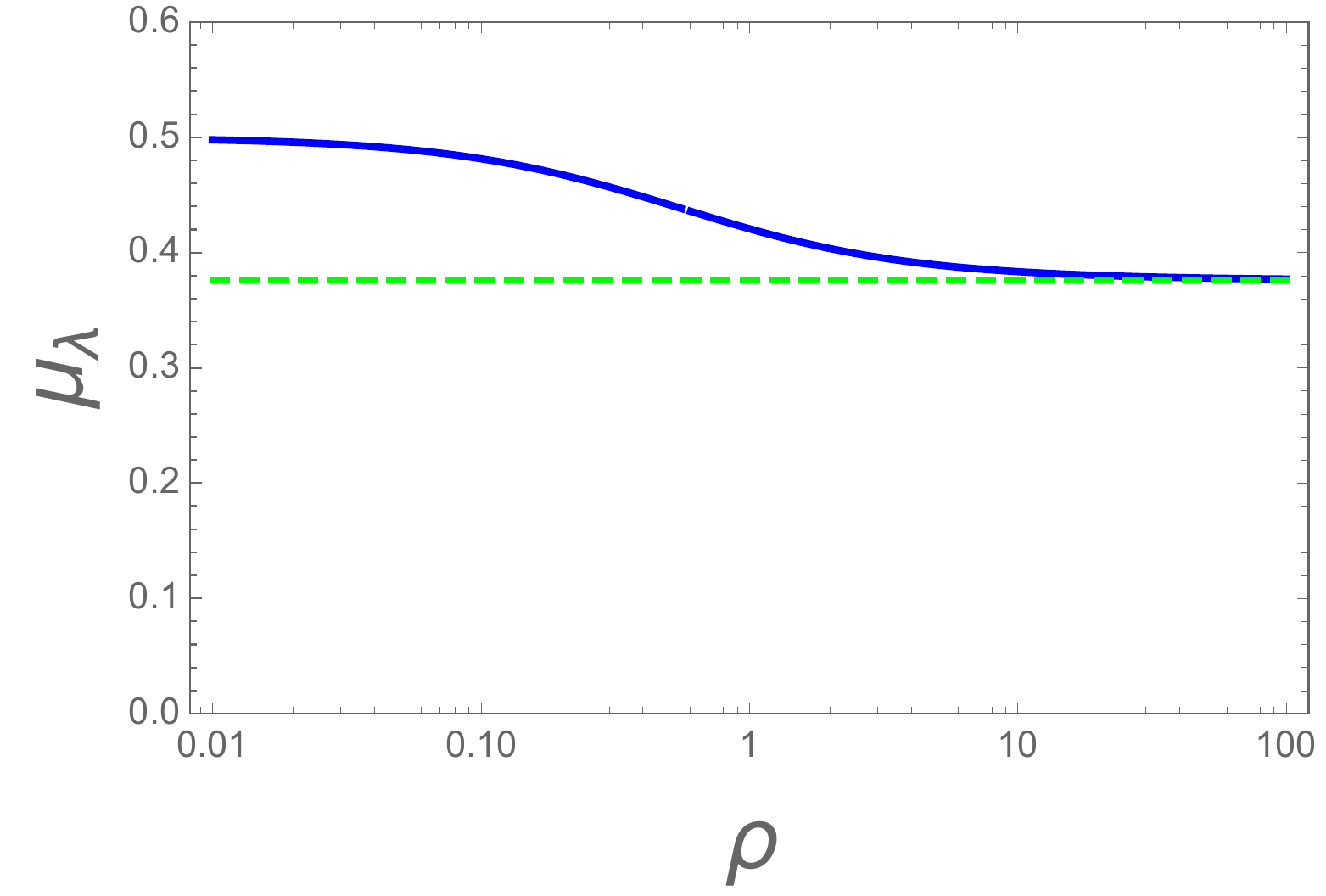}
\caption{The top panel: $ \mu_\lambda$ vs $\rho$. The lower panel: the same on a semi-log scale; the dotted line is the dirty limit value.
}
\label{f1}
\end{figure}

Following Ref.\,\onlinecite{HW}, one  defines the quantity
 \begin{equation}
 \mu_\lambda^{-1}= -\frac{T_c}{\lambda ^{-2}(0)}\left(\frac{d \lambda ^{-2}}{dT}\right )_{T_c}=-\left(\frac{d \rho_s}{dt}\right )_{t=1}\,,
 \label{eq8}
\end{equation}
where $t=T/T_c$, $\rho_s(t)=\lambda^2(0)/\lambda^2(t)$ is commonly called the superfluid density.  We then obtain from Eqs.\,(\ref{lambda-GL}), (\ref{lambda0}):
 \begin{eqnarray}
  \mu_\lambda^{-1}&=& \frac{16e^\gamma}{7\pi\zeta(3) \rho^2}
  \left[   \psi\left(\frac{1+\rho}{2}\right)-  \psi\left(\frac{1}{2}\right) -\frac{\pi^2\rho}{4}\right]\nonumber\\
&  \Big / &\left(1 +\frac{4\tan^{-1}\frac{ \eta-1 }{\sqrt{1-\eta^2} }}{\pi\sqrt{1-\eta^2}}\right),\qquad \eta=\rho\,e^\gamma.
  \label{HW1}
\end{eqnarray}
The top panel of Fig.\,\ref{f1} shows $ \mu_\lambda $ vs $\rho$. In the clean limit $ \mu_\lambda=0.5$, as it should. With increasing $\rho$,  $ \mu_\lambda $ decreases by about 30\%, but most of this change happens close to the clean limit where $0<\rho<1$.
The relative change of $ \mu_\lambda $  is about 9\% when $\rho$ varies in the realistic range from 1 to 10. Note also that for  scattering parameters $\rho\sim 100$ the dirty limit is reached with $ \mu_\lambda  = 7 \zeta(3)/4\pi e^\gamma \approx 0.376$. Curiously enough, the empirical two-fluid model with $\lambda^{-2}=\lambda^{-2}(0)(1-t^4)$ yields $ \mu_\lambda=0.25$.

\section{Relation between $\bm{\lambda^{-2}(0)}$ and    $\bm{(dH_{c2}/dT)_{T_c} }$  }


The slopes of  $H_{c2}$   at $T_c$ are given in  Ref.\,\onlinecite{HW} for isotropic type-II materials with arbitrary transport scattering:
\begin{eqnarray}
 &&    -\frac{dh}{dt}\Big|_{t=1}=3 \rho ^2 \left[ \psi\left( \frac{1}{2}\right)
- \psi\left(\frac{\rho +1}{2}\right) + \frac{\rho }{2} \psi^\prime \left(
 \frac{1}{2}\right)\right]^{-1},\qquad
 \label{h'_GL}
\end{eqnarray}
here, $t$ is the reduced temperature and
\begin{equation}
 h= H_{c2}\frac{\hbar^2v^2}{2\pi  T_c^2 \phi_0}\,.
\label{t,h}
\end{equation}
 Comparing this with the slope of $\lambda^{-2}$ near $T_c$, Eq.\,(\ref{lambda-GL}), one  sees that the $\rho$ dependences of these two quantities are inversed so that their product is  $\rho$-independent:
  \begin{equation}
\left(\frac{d \lambda ^{-2}}{dT}\,\frac{d H_{c2} }{dT}\right)_{T_c}=  \frac{128\pi^4
 }{7\zeta(3)\phi_0}\,  N(0)   .
 \label{new}
\end{equation}
Remarkably, the scattering parameter does not enter this relation at all. The only material parameter on the RHS is the density of states $N(0)$.
The physical reason for this result can be traced to the Rutgers thermodynamic relation,\cite{Rutgers} in which the product of slopes in Eq.\,(\ref{new}) is proportional to the transport-scattering-independent specific heat jump at $T_c$  (Anderson's theorem).

  In particular, this relation can be checked in the clean limit where at $T_c$
   \begin{equation}
\frac{d \lambda ^{-2}}{dT}= - \frac{16\pi e^2   N(0) v^2}{ 3c^2T_c}, \,\,\,
 \frac{d H_{c2} }{dT}  =  - \frac{24\pi  \phi_0  T_c
 }{7\zeta(3)\hbar^2v^2}.
 \label{clean}
\end{equation}
In the dirty limit, we have:
   \begin{equation}
  \frac{d \lambda ^{-2}}{dT}= - \frac{16\pi^3 e^2   N(0) v^2
 }{21\zeta( 3)c^2T_c\rho}, \,\,\,
 \frac{d H_{c2} }{dT} = - \frac{24\phi_0   T_c\rho
 }{ \pi v^2\hbar^2}.
 \label{dirty}
\end{equation}
The slopes of $\lambda ^{-2}$ at $T_c$ follow from Eq.\,(\ref{lambda-GL}); one can find  slopes of $H_{c2}$ in Ref.\,\onlinecite{HW}.

 Both clean and dirty limits are not quite realistic. According to Fig.\,\ref{f1}, within the  range $1<\rho<10$ one has approximate relations:
  \begin{equation}
 \left(\frac{d \lambda ^{-2}}{dT}\right )_{T_c}\approx - \,\frac{\lambda^{-2}(0)}{0.4T_c} \,,
 \label{lam'}
\end{equation}
while
  \begin{equation}
 \left(\frac{d H_{c2}}{dT}\right )_{T_c}\approx - \frac{H_{c2}(0)}{0.7 T_c} \,.
 \label{HW'}
\end{equation}

Using Eq.\,(\ref{new}) along with (\ref{lam'}) and (\ref{HW'}) one obtains:
  \begin{equation}
  H_{c2}(0)\,\lambda^{-2}(0) \approx 415  \,\frac{T_c^2N(0)}{\phi_0}\,.
 \label{Holam}
\end{equation}
This, however, holds  if there is no low-temperature   paramagnetic limiting of $H_{c2}$. Utilizing only Eq.\,(\ref{lam'}), one gets:
 \begin{equation}
-  \lambda^{-2}(0)  \left(\frac{d H_{c2}}{dT}\right )_{T_c} \approx 593  \,\frac{T_c N(0)}{\phi_0}\,,
 \label{lam0}
\end{equation}
a potentially useful relation, since the paramagnetism is not involved here and
one can express $\lambda^{-2}(0)$ in terms of the slope $(d H_{c2}/dT)_{T_c}$, $T_c$, and the density of states $N(0)$.
Noting that $N(0)=3\gamma/2\pi^2$ ($\gamma$ is the coefficient in linear low-$T$ dependence of the specific heat) and the slope $H_{c2}^\prime (T_c) $ are usually accessible, one can estimate a difficult to measure $ \lambda(0)$.
Note: in Eq.\,(\ref{lam0}), temperatures are given in energy units (erg); with temperature in Kelvins one has:
 \begin{eqnarray}
 - \lambda^{-2}(0)  \left(\frac{d H_{c2}}{dT}\right )_{T_c} \left(\frac{\rm G}{\rm cm^2\rm K}\right) \approx 593. \,\frac{k_B^2T_c N(0)}{\phi_0}\nonumber\\
   \approx  4.5 \times 10^8\left(\frac{1}{\rm cm^2\rm G}\right)T_c(\rm K)\,\gamma \left(\frac{{\rm erg}}{{\rm cm}^3{\rm K}^2 } \right).\qquad
 \label{Kelvins1}
\end{eqnarray}
This is perhaps the most useful result of our paper since it relates a difficult to measure $\lambda(0)$ to easily accessible slope $\left(d H_{c2}/dT\right )_{T_c}$.

\section{Discussion}

The above arguments   hold for isotropic s-wave materials with non-magnetic scattering. In the presence of pair-breaking or for other than s-wave order parameter and general Fermi surfaces, the Anderson theorem does not work, and  Eq.\,(\ref{new}) is not expected  to be valid.  Eqs.\,(\ref{HW}) may still hold, however with the factor $\mu_\lambda $ changing significantly with scattering, unlike the situation considered here. On the other hand, if the order parameter is constant at the Fermi surface of any shape (including multi-band structures), there is no obvious reason for our results to be inapplicable.

To show   how the obtained results can be applied for real materials, we estimate $  \lambda(0)$ of V$_3$Si and Nb$_3$Sn using  data of Orlando {\it et al}.\cite{Orlando}  For a sample of V$_3$Si with $T_c=16.4\,$K, $dH_{c2}/dT=-1.84\times 10^4\,$Oe/K, $\gamma=2.2\times 10^4\,$erg/cm$^3$K$^2$, Eq.\,(\ref{Kelvins1}) yields $  \lambda(0)\approx 106\,$nm.
Near $T_c$, $\lambda=\lambda_{GL}/\sqrt{1-t}$ and
Ref.\,\onlinecite{Orlando} provides $\lambda_{GL}=62\,$nm. In the clean limit we have $  \lambda(0)=\lambda_{GL}\sqrt{2}\approx 88\,$nm, a reasonably close  to 106\,nm given uncertain scattering parameters for these data (as shown below, the dirty limit assumption would give  $  \lambda(0)=1.63\lambda_{GL} \approx 102\,$nm).

For Nb$_3$Sn with $T_c=17.9\,$K, $dH_{c2}/dT=-1.83\times 10^4\,$Oe/K, $\gamma=1.1\times 10^4\,$erg/cm$^3$K$^2$, one obtains $  \lambda(0)\approx 144\,$nm, that corresponds to $\lambda_{GL}=\lambda(0)/\sqrt{2}=102\,$nm, whereas Ref.\,\onlinecite{Orlando} cites $\lambda_{GL}=64\,$nm (the dirty limit assumption would have given $ \lambda_{GL} \approx 88\,$nm).

  Another example is    Rh$_9$In$_4$S$_4$ with $T_c=2.25\,$K, $dH_{c2}/dT=-1.69\times 10^4\,$G/K, $\gamma=34\times 10^4\,$erg/mol\,K$^2$ = 0.21$\times 10^4\,$erg/cm$^3$K$^2$, and $\lambda_{GL}=575\,$nm. \cite{Udhara}  The  ratio of zero-$T$ BCS coherence length to the mean-free path for the sample studied was $\xi_0/\ell\sim 20-200$, i.e., it is  the dirty limit.  From Eq.\,(\ref{lambda-tensor1}) with   $\eta\gg 1$ one obtains the known expression for the dirty limit:
\begin{equation}
 \lambda^{-2}= \frac{8\pi^2 e^2 N(0)D}{c^2\hbar}\,  \Delta  \tanh\frac{\Delta }{ 2T}  \,,
 \label{lambda-dirty}
\end{equation}
($D=v\ell/3$ is the diffusivity). It is now readily shown that
\begin{equation}
\frac{ \lambda^{2}(0)}{ \lambda_{GL}^{2} }= \frac{4\pi^2 T_c}{7\zeta(3)\Delta_0
}\,.
 \label{lambda0/lamGL}
\end{equation}
Since, $\Delta_0/T_c\approx 1.76$, we estimate $ \lambda (0)\approx 1.63  \lambda_{GL}\approx 939\,$nm.

On the other hand,   Eq.\,(\ref{Kelvins1}) yields $ \lambda (0)\approx   892\,$nm. A reasonable agreement between our model and the data of this case might be due to the fact that the strong scattering washes away anisotropies of the order parameter thus making the material ``more BCS-like". Besides, the strong scattering excludes possibility of other than s-wave symmetry since even the transport scattering for a non-s-wave symmetry is pair breaking and superconductivity disappears well before the dirty limit is reached.

The multi-band MgB$_2$ is an example for which our model should not work. Still, taking $T_c\approx 39\,$K, $dH_{c2}/dT=-0.41\times 10^4\,$G/K as given in Ref.\,\onlinecite{BCK} along with $\gamma=7.2\times 10^2 \,$erg/cm$^3$K$^2$ as provided by Ref.\,\onlinecite{Junod}, with the help of Eq.\,(\ref{Kelvins1}) we estimate $ \lambda (0)\approx   176\,$nm.   Ref.\,\onlinecite{Junod} cites 185\,nm obtained from thermodynamic data, and a close value of 180\,nm reported from analysis of microwave response.\cite{MgB2} One may say that proximity of these numbers does not mean much. On the other hand, it shows that formulas based on the penetration depth property represented by Eq.\,(\ref{HW}), are quite robust and can be used for rough estimates in variety of situations, similar to what is commonly done with the $H_{c2}$ property of Eq.\,(\ref{HW0}).

Still, since the above derivation of $\lambda(0)$, Eq.\,(\ref{lambda0}), has been done for s-wave superconductors with only transport scattering, one should not expect consequences of this equation to hold in the presence of pair breaking, be it due to the spin-flip or to other than s-wave order parameter. To check this we turn to well-studied CeCoIn$_5$, a clean superconductor with $T_c=2.3\,$K and $\gamma=3\times 10^4\,$erg/cm$^3$K$^2$.\cite{Movs} The penetration depth in this material  turned out\cite{KPPetr} to satisfy with high accuracy the  relation $\lambda= \lambda(0) /\sqrt{1-t^2}$ established by Abrikosov-Gor'kov  for a strong pair breaking.\cite{AG} The fit to the data gives $\lambda(0)=358\,$nm and $dH_{c2}/dT=-11.5\times 10^4\,$G/K.\cite{KPPetr}
 We thus have all information needed to calculate $\lambda(0)$ with the help of Eq.\,(\ref{Kelvins1}) that gives $\lambda(0)=608\,$nm.
 The discrepancy is larger yet,  if we compare with $\lambda(0) =196\,$nm , as
found in microwave measurements.\cite{Truncik}
  Clearly, our model fails in this case indicating the pair breaking as a possible culprit.

 Also, our model fails being applied to KFe$_2$As$_2$ with $T_c=3.5\,$K, $\gamma=1.53\times 10^4\,$erg/cm$^3$K$^2$ and $dH_{c2,c}/dT=-0.6\,$T/K for the field along the $c$ direction.\cite{Abdel-Hafiez,Hardy,Yong} Equation (\ref{Kelvins1}) of the model yields $\lambda(0)=158\,$nm while the literature values are close to 200\,nm.\cite{Ohishi,Abdel-Hafiez,Hyunsoo,Hyunsoo1} One of the reasons for disagreement is possible d-wave order parameter   and a relatively strong anisotropy of superconducting properties.

In conclusion, we consider isotropic s-wave superconductors without pair-breaking scattering, but with arbitrary potential scattering. We establish  new relations of the London penetration depth at $T = 0$ with its variation at $T_c$ and
with the slope of the upper critical field  $H_{c2}$. This gives
a relatively simple way to estimate difficult to measure
$\lambda(0)$ from measured $H_{c2}$ near $T_c$ and the specific heat (needed to estimate the density of states $N(0)$ at the Fermi level). On
the other hand, if the obtained estimate comes out unreasonable or very different from independent measurement, this may signal the unconventional superconductivity or pair breaking in the studied material. We, therefore, believe that
our work provides a useful practical tool for researchers dealing with experimental superconductivity.
 \\

We  thank S. Bud'ko  for discussions and  comments.
The work was supported by the U.S. Department of Energy (DOE), Office of Science, Basic Energy Sciences, Materials Science and Engineering Division.  Ames Laboratory  is operated for the U.S. DOE by Iowa State University under contract \# DE-AC02-07CH11358.


\references

\bibitem{HW}E. Helfand and N. R. Werthamer, Phys. Rev. {\bf 147}, 288 (1966).

 \bibitem{3authors}A. A. Abrikosov, L. P. Gor'kov, I. E. Dzyaloshinskii, {\it
Methods of Quantum Field Theory in Statistical Physics}, Englewood Cliffs, N.J., Prentice-Hall, 1963.

\bibitem{nonloc} V.  G. Kogan, A. Gurevich, J. H. Cho, D. C. Johnston,   Ming Xu, J. R. Thompson, A. Martynovich, \prb {\bf 54}, 12386 (1996).

\bibitem{PK-ROPP} R. Prozorov and V. G. Kogan, Rep. Prog. Phys. {\bf 74}, 124505 (2011).

\bibitem {Homes} V. G. Kogan,   \prb {\bf  87}, 220507(R) (2013).

 \bibitem{Rutgers} H. Kim,  V. G. Kogan,  K. Cho,  M. A. Tanatar,  and R. Prozorov, \prb {\bf 87}, 214518 (2013).

\bibitem{KP} V. G. Kogan  and R. Prozorov, \prb {\bf 88}, 024503 (2013).


\bibitem{Orlando} T. P. Orlando, E. J. McNiff Jr., S. Foner, and M.~R.~Beasley, \prb {\bf 19}, 4545 (1979).

\bibitem{Udhara} U. S. Kaluarachchi,  Qisheng Lin,  Weiwei Xiey,  V.
Taufour,  S. L. Bud'ko,  G. J. Miller,  and P. C. Canfield, \prb {\bf 93}, 094524 (2016).

\bibitem{BCK} S. L. Bud'ko, P. C. Canfield, and V. G. Kogan, Physica C, {\bf 382}, 85 (2002).

\bibitem{Junod}  A. Junod, Y. Wang, F. Bouquet, P. Toulemonde, in {\it Studies of High Temperature Superconductors}, 38, ed. by A. V. Narlikar, Nova Science, New York (2001), p. 179.

\bibitem{MgB2}B. P. Xiao, X. Zhao, J. Spradlin, C. E. Reece, M. J. Kelley, T. Tan,  and X. X. Xi, Supercond. Sci.  Technol.  {\bf 25},  095006 (2012).

\bibitem{Movs}  R. Movshovich,  M. Jaime,  J. D. Thompson,  C. Petrovic,  Z. Fisk,  P. G. Pagliuso,  and J. L. Sarrao, \prl   {\bf 86} 5152 (2001).

\bibitem{KPPetr} V. G. Kogan, R Prozorov, and C Petrovic, J. Phys.: Condens. Matter,  {\bf 21}, 1 (2009).


\bibitem{AG}A. A. Abrikosov and L. P. Gor'kov, Zh. Eksp. Teor. Fiz. {\bf 39}, 1781 (1960) [Sov. Phys. JETP, {\bf  12}, 1243 (1961)].

\bibitem{Truncik}C. J. S. Truncik, W. A. Huttema, P. J. Turner, S. Ozcan,
N. C. Murphy, P. R. Carriere, E. Thewalt, K. J. Morse,
A. J. Koenig, J. L. Sarrao, and D. M. Broun, Nat. Comm.
{\bf 4}, 2477 (2013).

\bibitem{Abdel-Hafiez} M. Abdel-Hafiez,  S. Aswartham, S. Wurmehl, V. Grinenko,
S. L. Drechsler, S. Johnston, A.U.B. Wolter-Giraud, and B. B\"{u}chner,
arXiv:1110.6357 (2012).

\bibitem{Hardy}F. Hardy, A. E. B\"{o}hmer, D. Aoki, P. Burger, T. Wolf, P.
Schweiss, R. Heid, P. Adelmann, Y. X. Yao, G. Kotliar, J.
Schmalian, and C. Meingast Phys. Rev. Lett. {\bf 111}, 027002
(2013).

\bibitem{Yong} Yong Liu, M. A. Tanatar, V. G. Kogan, Hyunsoo Kim,
T. A. Lograsso, and R. Prozorov Phys. Rev. B {\bf 87}, 134513
(2013).

\bibitem{Ohishi} K. Ohishi, Y. Ishii, I. Watanabe, H. Fukazawa, T. Saito, Y. Kohori, K. Kihou, Chul-Ho Lee, H. Kito, A. Iyo,   H.~Eisaki,  Journal of Physics: Conference Series {\bf 400},  022087 (2012).

\bibitem{Hyunsoo} R. T. Gordon, H. Kim, N. Salovich, R. W. Giannetta, R.~M.~Fernandes, V. G. Kogan, T. Prozorov, S. L. Bud'ko, P.~C.~Canfield, M. A. Tanatar, R.  Prozorov, \prb {\bf 82}, 054507 (2010).

\bibitem{Hyunsoo1}H. Kim, M. A. Tanatar, Yong Liu, Zachary Cole Sims,
Chenglin Zhang, Pengcheng Dai, T. A. Lograsso, and R.
Prozorov, Phys. Rev. B 89, 174519 (2014).
\end{document}